\documentclass[fleqn,usenatbib]{mnras}

\usepackage{newtxtext,newtxmath}
\usepackage[T1]{fontenc}

\DeclareRobustCommand{\VAN}[3]{#2}
\let\VANthebibliography\thebibliography
\def\thebibliography{\DeclareRobustCommand{\VAN}[3]{##3}\VANthebibliography}

\usepackage{graphicx}	% Including figure files
\usepackage{amsmath}	% Advanced maths commands
\usepackage{hyperref}

\title[The lithium evolution in the Galactic disc]{The temporal and spatial variations of lithium abundance in the Galactic disc}

\author[Tiancheng Sun et al.]{
Tiancheng Sun,$^{1,2,7}$\thanks{E-mail: isaac@mail.bnu.edu.cn}
Shaolan Bi,$^{1,2}$\thanks{E-mail: bisl@bnu.edu.cn}
Xunzhou Chen,$^{3}$\thanks{E-mail: cxz@zhejianglab.org}
Yuxi (Lucy) Lu,$^{4,5,6}$
Yuqin Chen,$^{7,1}$
Ming-Yi Ding,$^{7,8}$
\newauthor
Jianrong Shi,$^{7,8}$
Hongliang Yan,$^{7,8,1}$
and Zhishuai Ge$^{9}$
\\
$^{1}$Institute for Frontiers in Astronomy and Astrophysics, Beijing Normal University,  Beijing 102206, China\\
$^{2}$School of Physics and Astronomy, Beijing Normal University, Beijing 100875, People’s Republic of China\\
$^{3}$Research Center for Astronomical Computing, Zhejiang Laboratory, Hangzhou 311100, China\\
$^{4}$American Museum of Natural History, Central Park West, Manhattan, NY, USA\\
$^{5}$Department of Astronomy, The Ohio State University, Columbus, 140 W 18th Ave, OH 43210, USA\\
$^{6}$Center for Cosmology and Astroparticle Physics (CCAPP), The Ohio State University, 191 W. Woodruff Ave., Columbus, OH 43210, USA\\
$^{7}$CAS Key Laboratory of Optical Astronomy, National Astronomical Observatories, Chinese Academy of Sciences, Beijing 100101, China\\
$^{8}$School of Astronomy and Space Science, University of Chinese Academy of Sciences, Beijing 100049, China\\
$^{9}$Beijing Planetarium, Beijing Academy of Science and Technology, Beijing, 100044, China\\
}

\date{Accepted 2024 November 19. Received 2024 November 7; in original form 2024 June 6}

\pubyear{\the\year{}}

\begin{document}
\label{firstpage}
\pagerange{\pageref{firstpage}--\pageref{lastpage}}
\maketitle

% Abstract of the paper
\begin{abstract}
This study investigates the temporal and spatial variations in lithium abundance within the Milky Way using a sample of 22,034 main-sequence turn-off (MSTO) stars and subgiants, characterised by precise stellar ages, 3D NLTE (non-local thermodynamic equilibrium) lithium abundances, and birth radii. Our results reveal a complex variation in lithium abundance with stellar age: a gradual increase from 14 Gyr to 6 Gyr, followed by a decline between 6 Gyr and 4.5 Gyr, and a rapid increase thereafter.
We find that young Li-rich stars (ages $<$ 4 Gyr, A(Li) $>$ 2.7 dex) predominantly originate from the outer disc. By binning the sample according to guiding center radius and z$_{\rm max}$, we observe that these young Li-rich stars migrate radially to the local and inner discs. In addition, the stars originating from the inner disc experienced a rapid Li enrichment process between 8 Gyr and 6 Gyr.
Our analysis suggests that the age range of Li-dip stars is 4-5 Gyr, encompassing evolution stages from MSTO stars to subgiants. The Galactic radial profile of A(Li) (with respect to birth radius), as a function of age, reveals three distinct periods: 14-6 Gyr ago, 6-4 Gyr ago, and 4-1 Gyr ago. Initially, the lithium abundance gradient is positive, indicating increasing Li abundance with birth radius. During the second period, it transitions to a negative and broken gradient, mainly affected by Li-dip stars. In the final period, the gradient reverts to a positive trend.
\end{abstract}

% Select between one and six entries from the list of approved keywords.
% Don't make up new ones.

%%%%%%%%%%%%%%%%%%%%%%%%%%%%%%%%%%%%%%%%%%%%%%%%%%

%%%%%%%%%%%%%%%%% BODY OF PAPER %%%%%%%%%%%%%%%%%%

\section{Introduction}

The evolution of lithium (Li) is a crucial question in modern astrophysics, especially its chemical evolution history in the Milky Way. According to the Standard Big Bang Nucleosynthesis model (SBBN), the primordial Li abundance is predicted to be A(Li)\footnote{A(Li) = log[N(Li)/N(H) + 12], where N(Li) and N(H) are the number densities of lithium and hydrogen, respectively.} $\sim$ 2.7 dex \citep{2003PhLB..567..227C,2008JCAP...11..012C,2016RvMP...88a5004C,2003ApJS..148..175S,2012ApJ...744..158C,2014AIPC.1594...12C,2014JCAP...10..050C,2018PhR...754....1P,2020JCAP...03..010F}, which is a factor of 3--4 higher than measured in halo dwarf stars \citep[A(Li) = 2.05--2.2 dex,][]{1982A&A...115..357S,1997MNRAS.285..847B,2006ApJ...644..229A,2007A&A...462..851B}. This discrepancy is the well-known cosmological lithium problem \citep{2012MSAIS..22....9S}.
On the other hand, the discovery of unevolved stars with high Li abundances \citep[A(Li) = 3.26 dex,][]{2009LanB...4B..712L} signifies an imprint of Li-enrichment within the interstellar medium (ISM) during the Galaxy lifetime.
Furthermore, the release of high-resolution spectroscopic data has revealed distinct Li evolution patterns among stars in the thin and thick discs. 
The thin disc stars exhibit an increasing Li abundance with higher metallicity \citep[e.g.,][]{2016A&A...595A..18G,2018A&A...610A..38F,2018A&A...615A.151B}.
However, the relation between Li abundance and metallicity in thick disc stars is less clear. Some studies report an increase in Li abundance with metallicity \citep{2016A&A...595A..18G,2018A&A...610A..38F}, while others suggest a decrease \citep{2015A&A...576A..69D,2018A&A...615A.151B}, or a flat trend \citep{2012ApJ...756...46R}. These discrepancies arise from the methods used to select thick disc stars, as discussed in \cite{2018A&A...615A.151B}. They suggest that selecting thick disc stars based on stellar age significantly reduces contamination from thin disc stars.

A multitude of nucleosynthesis processes and sources have been proposed to explain the increase of Li abundance in disc stars, including spallation by Galactic cosmic ray spallation \citep[GCC,][]{1970Natur.226..727R} and specific stellar evolution phases: core-collapse supernovae \citep[CCSN,][]{1990ApJ...356..272W,2019ApJ...872..164K}, novae \citep{1975A&A....42...55A,1996ApJ...465L..27H,2002AIPC..637..104J,2015Natur.518..381T}, low-mass giants \citep{1999ApJ...510..217S}, and the asymptotic giant branch stars (AGB) \citep{1992ApJ...392L..71S,1999A&A...351..273A}.
Among the various sources, novae are the primary contributors of Li in the Galaxy \citep{1991A&A...248...62D,1999A&A...352..117R,2001A&A...374..646R,2019MNRAS.489.3539G,2021A&A...653A..72R}. Significant contributions also come from AGB stars \citep{1991A&A...248...62D} and GCC \citep{1999A&A...352..117R,2012A&A...542A..67P}, with minor contributions from Li-rich giant stars \citep{2001A&A...374..646R}. The contribution of CCSN to Li production remains uncertain \citep{1999A&A...352..117R,2012A&A...542A..67P,2021A&A...653A..72R}.

Despite the discernment of multiple production sites, the constraints on stellar yields remain inadequate. For instance, observations indicate a potential decrease in Li abundance among cool field main sequence (MS) stars at super-solar metallicities \citep{2015A&A...576A..69D,2016A&A...595A..18G,2020A&A...634A.130B,2020AJ....159...90S}.
Several studies propose a correlation between the observed decrease in Li abundance and radial migration \citep{2019A&A...623A..99G,2022A&A...668L...7D,2023MNRAS.520.4815Z}. Specifically, older and Li-depleted stars born in the inner regions of the Galactic disc migrated to the solar vicinity. This migration renders the upper envelope of the A(Li) versus [Fe/H] diagram ineffective for accurately tracing Li evolution, particularly at the metal-rich end \citep[e.g., ][]{2019A&A...623A..99G}.
To delve into the Li evolution in the Milky Way, a prerequisite is a statistically robust and homogeneous sample with precise age estimates and birth radii.

Another problem of Li depletion, Li-dip, was initially observed in main-sequence stars in the Hyades open cluster \citep{1965ApJ...141..610W,1986ApJ...302L..49B,2000A&A...354..216B,2016ApJ...830...49B}, reveal that F-dwarfs within a 300 K temperature range around 6600 K (1.0-1.5 M$_{\odot}$) exhibit a depletion in A(Li) of over 2 dex compared with stars out of this region, forming a distinct 'Li dip'. This phenomenon is confirmed to exist in field stars \citep{1999A&A...348..487R,2001A&A...371..943C,2004MNRAS.349..757L,2012ApJ...756...46R,2018A&A...614A..55A,2018A&A...615A.151B,2020MNRAS.497L..30G} and clusters like M34 \citep{1997AJ....114..352J}, NGC 752 \citep{1986ApJ...309L..17H}, and M67 \citep{1995ApJ...446..203B}, but not in youngest clusters with ages less than about 100 Myr like the Pleiades \citep{1988ApJ...327..389B}, indicating it develops later in the main sequence phase \citep{2005A&A...442..615S}.
Despite current models attributing the Li dip to atomic diffusion \citep{1986ApJ...302..650M}, rotational mixing \citep{1992ApJS...78..179P,2010A&A...519L...2E}, and internal gravity waves \citep{2000A&A...354..943M,2005Sci...309.2189C}, several studies have demonstrated a correlation between the mass of Li-dip star and the metallicity of cluster \citep[e.g.,][]{1995ApJ...446..203B,2009AJ....138.1171A}. This correlation implies that any comprehensive explanation for the Li dip must consider not only its morphology but also the potential relations among stellar mass, age, and metallicity \cite[e.g.,][]{2015A&A...576A..69D}. To date, a consistent explanation for the Li dip remains elusive, highlighting the need for accurate Li abundance measurements across diverse stars to understand these mechanisms' dependence on stellar properties.

As the first 3D NLTE (non-local thermodynamic equilibrium) analysis of Li applied to a large spectroscopic survey, \cite{2024MNRAS.528.5394W} have published 3D NLTE Li abundances for 581,149 stars observed in GALAH DR3. For the first time, the evolution of Li-dip stars beyond the main sequence turn-off and along the subgiant branch has been traced. In addition, based on spectroscopic data from GALAH DR3 and astrometric data from GAIA DR3, \cite{2023MNRAS.523.1199S} (hereafter Paper I) and \cite{2023arXiv231105815S} (hereafter Paper II) have obtained precise ages for approximately 45,000 subgiants and turnoff stars, with median age uncertainties better than 10\%.
The integration of the 3D NLTE Li abundance with these precise ages allows for an unprecedented characterisation of the temporal evolution of Li abundance.

This paper is organised as follows: Section \ref{sec:data} outlines the sample selection, age estimation, and calculation of the birth radius R$_{\rm birth}$. Section \ref{sec:result} presents the temporal and spatial variations in Li abundances. The results are summarised in Section \ref{sec: conclu}.

\section{Data and method} \label{sec:data}

This work is based on the spectroscopic data from the Third Data Release of the Galactic Archaeology with HERMES survey \citep[GALAH DR3,][]{2021MNRAS.506..150B}, including stellar parameters ($T_{\rm eff}$, $\log g$, [Fe/H], $V_{mic}$, $V_{broad}$, $V_{rad}$) and up to 30 elemental abundances for 588,571 stars. The derivation of stellar parameters is based on optical spectra with a typical resolution of R $\sim$ 28,000. 
% The [Fe/H] from the GALAH DR3 catalogue \citep{2021MNRAS.506..150B} are calculated using the 1D NLTE methodology \citep{2020A&A...642A..62A}.
In addition, the Li abundance employed in this study were derived by \citet{2024MNRAS.528.5394W} using 3D hydrodynamic model atmospheres with NLTE radiative transfer. This 3D NLTE Li abundance catalogue is available on the GALAH DR3 website \footnote{\url{https://cloud.datacentral.org.au/teamdata/GALAH/public/GALAH_DR3/}}.

\begin{figure}
% \plotone{Figure2/3D_Li_rg_full.png}
\includegraphics[scale=1.1]{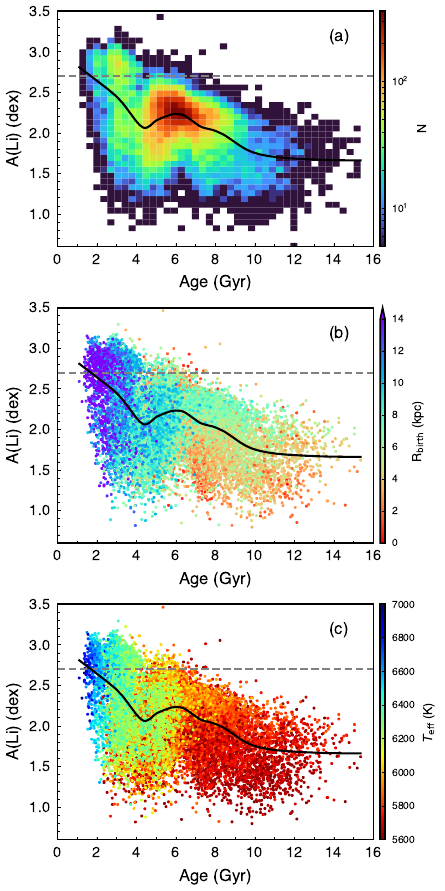}
\caption{Age-A(Li) distributions, colour-coded by the stellar number density (a), R$_{\rm birth}$ (b) and $T_{\rm eff}$ (c). The black lines represent the fitting results by local non-parametric regression.
The fitting procedure is performed using the Locally Weighted Scatterplot Smoothing (LOESS) model with the following parameters: a smoothing fraction (frac) of 0.15, 20 iterations (it), and no weighting function adjustment (delta = 0).
The grey dashed lines represent the typical SBBN Li abundances of A(Li) = 2.7 dex.
To clearly illustrate the distribution of the R$_{\rm birth}$ for the sample stars in panel (b), Fig.~\ref{full_rb_mean} in the Appendix \ref{appendix} presents the mean R$_{\rm birth}$ binned in age and A(Li).
\label{fig1}}
\end{figure}

\begin{figure}
% \plotone{Figure2/3D_Li_rg_full.png}
\includegraphics[scale=0.67]{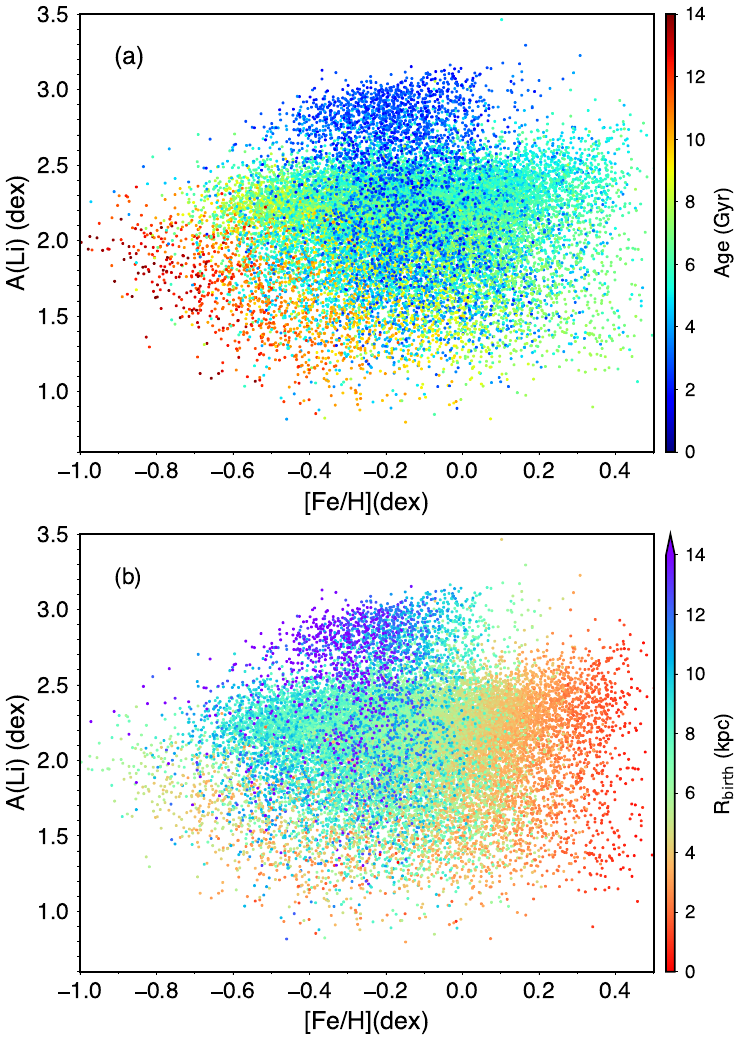}
\caption{[Fe/H]-A(Li) distributions, colour-coded by the stellar age (a) and R$_{\rm birth}$ (b). The [Fe/H]-A(Li) distribution of old stars with age $>$ 8 Gyr is shown in Fig.~\ref{fig:feh-ali-old} in the Appendix \ref{appendix2}.
\label{fig:feh-ali}}
\end{figure}

\begin{figure}
% \plotone{Figure2/3D_Li_rg_full.png}
\includegraphics[scale=0.67]{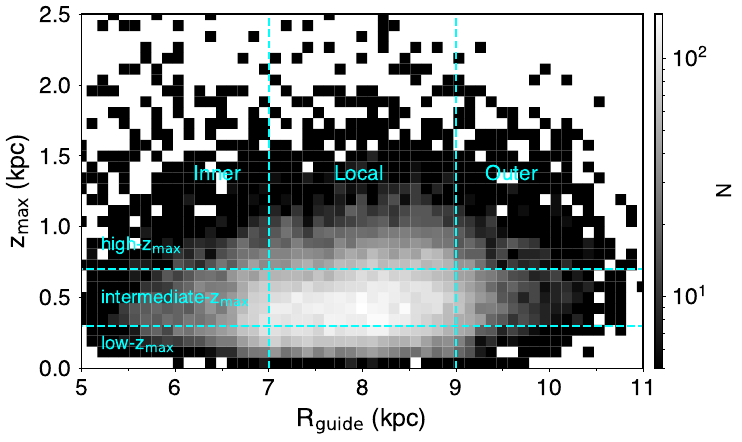}
\caption{R$_{\rm guide}$-z$_{\rm max}$ distribution of sample stars, colour-coded by the stellar number density.
The vertical dashed lines indicate the division into three R$_{\rm guide}$ bins (inner, local, and outer) at
R$_{\rm guide}$ = 7 and 9 kpc. The horizontal dashed lines indicates the division of each R$_{\rm guide}$ bin into three z$_{\rm max}$ bins (high-z$_{\rm max}$, intermediate-z$_{\rm max}$, low-z$_{\rm max}$) at z$_{\rm max}$ = 0.3 and 0.7 kpc.
\label{fig:rg-zmax}}
\end{figure}

The star sample utilised in this work comes from Paper II, which presents a sample of the main-sequence turnoff (MSTO) and subgiant stars, with a median relative age uncertainty of 9.8\% across the age range of 1–13.8 Gyr. 
Paper II has excluded stars with significant model systematic bias, whose inferred ages are 2-sigma larger\footnote{For a certain star, age $-$ 2*age$\_$uncertainty $>$ 13.8 Gyr.} than the age of the universe \citep[13.8 Gyr,][]{2016A&A...594A..13P}.
Binaries systems identified by \cite{2020A&A...638A.145T} and \cite{2023ApJS..264...41Y} have been excluded in this sample. Additionally, a criterion based on Gaia DR3 parameters is applied, selecting stars with a Gaia re-normalised unit weight error (RUWE) of less than 1.2 \citep{2022MNRAS.510.2597R}.

\subsection{Stellar age determination}

Stellar ages from Paper II are estimated by matching the Gaia luminosity \citep{2023ApJS..264...41Y}, GALAH spectroscopic stellar parameters ($T_{\rm eff}$, [Fe/H], [$\alpha$/Fe], and [O/Fe]), with Oxygen-enhanced stellar models, using a Bayesian approach \citep{2010ApJ...710.1596B}. 
These models employ an individual enhancement factor for oxygen, allowing for the independent specification of oxygen abundance (refer to Paper II for further details). The enhancement factors for the other $\alpha$-elements, including Ne, Mg, Si, S, Ca, and Ti, are maintained with the same enhancement factor ([$\alpha$/Fe] = 0, 0.1, 0.2, 0.3). The methods used for age determination of each star in Paper II are outlined as follows:

For each star, we utilise a set of stellar models characterised by the corresponding [$\alpha$/Fe] and [O/Fe] values, specifically selecting those models whose [$\alpha$/Fe] and [O/Fe] values are closest to the observed values of the star. To determine the most probable stellar models from evolution tracks, we apply a 3-sigma error (i.e., three times the observational error) to define the likelihood that matches the observed constraints ($T_{\rm eff}$, [Fe/H], luminosity).

Following the fitting method introduced by \cite{2010ApJ...710.1596B}, we conduct a comparison between the model predictions and their corresponding observational properties $D$. This comparison allows us to calculate the overall probability of the model $M_i$ with posterior probability $I$,
\begin{equation}\label{e1}
p\left(M_{i}\mid D,I\right)=\frac{p\left(M_{i}\mid I\right) p\left(D\mid M_{i}, I\right)}{p(D\mid I)}.
\end{equation}
with $p$($M_i$ $\mid$ $I$) representing the uniform prior probability for a specific model, and $p$(D $\mid$ $M_i$, $I$) denoting the likelihood function: 
\begin{equation}\label{e2}
\begin{aligned}
p\left(D\mid M_{i},I\right)=L(T_{eff},[Fe/H],lum)\\
=L_{T_{eff}}L_{[Fe/H]}L_{lum},
\end{aligned}
\end{equation}
where the likelihood function is given by
\begin{equation}
L = \frac{1}{\sqrt{2\pi}\sigma} \exp\left(\frac{-\chi^2}{2}\right),
\end{equation}
and
\begin{equation}
\chi^2 = \left(\frac{\phi_{\text{obs}} - \phi_{\text{model}}}{\sigma}\right)^2.
\end{equation}
Here $\sigma$ is the error of the observation $\phi_{\text{obs}}$. The $p$($D$ $\mid$ $I$) in Equation \ref{e1} is a normalisation factor for the specific model probability:
\begin{equation}\label{e4}
p(D \mid I)=\sum_{j=1}^{N_{m}} p\left(M_{j} \mid I\right) p\left(D \mid M_{j}, I\right).
\end{equation}
where $N_m$ is the total number of selected models. The uniform priors $p$($M_i$ $\mid$ $I$) can be cancelled, giving the simplified Equation (1) as :
\begin{equation}\label{e5}
p\left(M_{i} \mid D, I\right)=\frac{p\left(D \mid M_{i}, I\right)}{\sum_{j=1}^{N_{m}} p\left(D \mid M_{j}, I\right)}.
\end{equation}
We derive the probability distribution for each star with Equation \ref{e5} and fit a Gaussian function to the resulting likelihood distribution. The mean and standard deviation of the Gaussian profile serve as estimates and uncertainties, respectively.

\begin{figure*}
\includegraphics[scale=1]{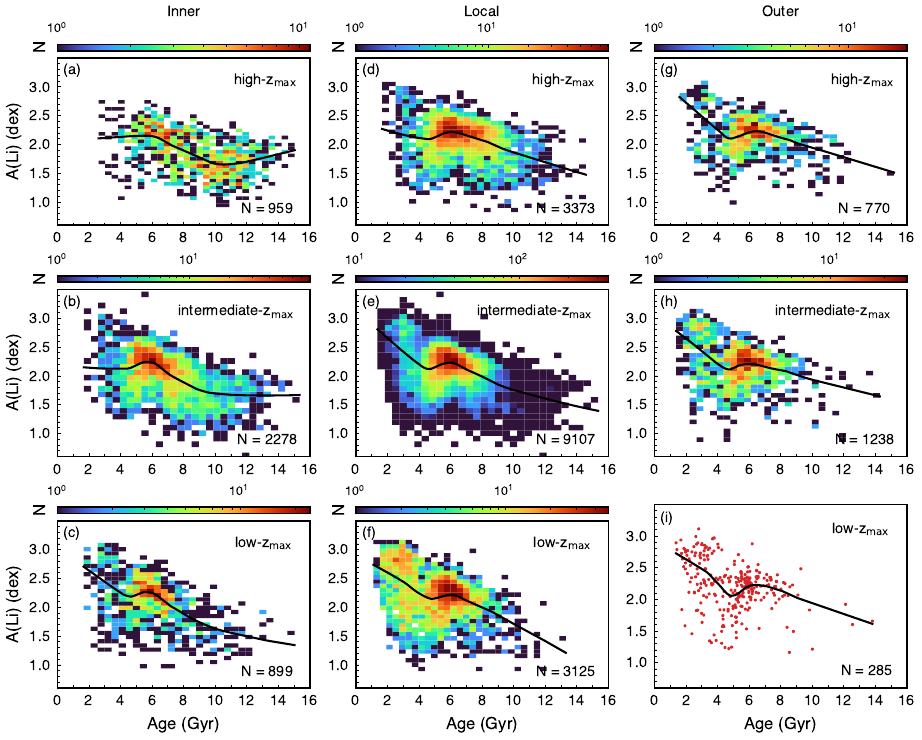}
\caption{Age-A(Li) distributions in nine different bins of R$_{\rm guide}$ and z$_{\rm max}$. (a-h): Colour-coded by the stellar number density. (i): Red dots represent the stars in the outer disc at low-z$_{\rm max}$ region. The numbers of stars in each bin are shown in the bottom-right corner of each panel. 
The panels are arranged according to the division in Fig.~\ref{fig:rg-zmax}.
The black lines represent the fitting result by local nonparametric regression. 
The fitting procedure is performed using the Locally Weighted Scatterplot Smoothing (LOESS) model with the following parameters: a smoothing fraction (frac) of 0.4, 20 iterations (it), and no weighting function adjustment (delta = 0).\label{9fig}}
\end{figure*}

\subsection{Target selection and the calculation of birth radius}

To avoid significant Li destruction in the coolest dwarf stars, we mainly target the MSTO and subgiant stars with $T_{\rm eff}$ ranging from 5600 K to 7000 K, and $\log g$ ranging from 3.4 to 4.1. The resulting sample comprises 22,034 field stars with 3D NLTE Li measurements (flag$\_$Ali$\_$DR3 $<$ 2). Although our sample includes a small number of early subgiant stars in addition to MSTO stars, these subgiant stars exhibit no significant reduction in surface A(Li) attributable to the first dredge-up \citep[e.g.,][]{1999ApJ...510..232B}. As illustrated in Fig.~\ref{logg_ali} in the Appendix \ref{appendix}, we observe no pronounced decrease in Li abundance with respect to $\log g$.

To calculate the birth radius R$_{\rm birth}$, we adopt the methodology outlined in \cite{2022arXiv221204515L}, which enhances the original approach introduced in \cite{2018MNRAS.481.1645M}. \cite{2022arXiv221204515L} establishes a linear relation between the metallicity gradient of the interstellar medium (ISM) at a given lookback time (\mbox{$\rm \nabla [Fe/H](\tau)$}), and the metallicity range observed in stars of the same age (\mbox{$\text{Range}\widetilde{\mbox{$\rm [Fe/H]$}}(age)$}), using two different sets of cosmological simulations. Consequently, the birth radius (R$_{\rm birth}$) of an individual star, assuming a linear gradient of metallicity in the ISM, is mathematically expressed as a function of its metallicity and age:
\begin{equation} \label{eqn:rb}
\rm R_{b}(age, [Fe/H]) = \frac{[Fe/H] - [Fe/H](0, \tau)}{\nabla [Fe/H](\tau)}.
\end{equation}
In which \mbox{$\rm [Fe/H](0, \tau$)} and \mbox{$\rm \nabla [Fe/H](\tau)$} are taken from Table A1 of \cite{2022arXiv221204515L} with linear interpolation in between points.

\begin{figure*}
\includegraphics[scale=1]{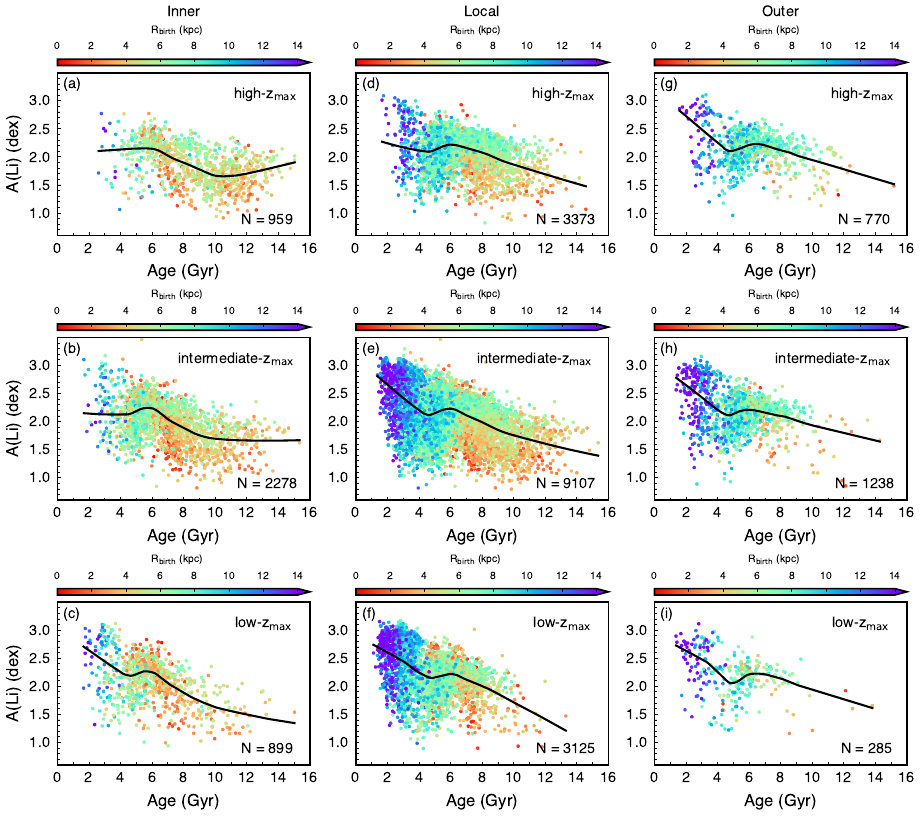}
\caption{Age-A(Li) distributions in nine different bins of R$_{\rm guide}$ and z$_{\rm max}$, color-coded by the R$_{\rm birth}$. 
The panels are arranged according to the division in Fig.~\ref{fig:rg-zmax}. The numbers of stars in each bin are shown in the bottom-right corner of each panel. The black lines represent the fitting result by local nonparametric regression. The fitting procedure is performed using the Locally Weighted Scatterplot Smoothing (LOESS) model with the following parameters: a smoothing fraction (frac) of 0.4, 20 iterations (it), and no weighting function adjustment (delta = 0).
\label{9fig-color}}
\end{figure*}

\section{Result} \label{sec:result}

\subsection{Age-A(Li) and [Fe/H]-A(Li) distributions} \label{sec:4.1}

Fig.~\ref{fig1} presents the age-A(Li) distributions of sample stars. 
We adopt local non-parametric regression on the data in each panel using the PYTHON package STATSMODELS \citep{seabold2010statsmodels}, specifically the Locally Weighted Scatterplot Smoothing (LOESS) model. This approach is well-suited for capturing non-linear trends in data by applying local linear regression across small subsets of the data. The weight assigned to each point during the fitting process is determined by its proximity to the target point, with closer points receiving higher weights.
As shown in Fig.~\ref{fig1}(a), the Li abundance (A(Li)) exhibits a complex temporal variation. Initially, A(Li) increases from $\sim$1.5 dex at 14 Gyr to $\sim$2.2 dex at 6 Gyr. 
Subsequently, it declines slightly to around 2.1 dex at 4.5 Gyr. Following this dip, the Li abundance rises again, surpassing 2.7 dex at 2 Gyr. Based on the birth radius of the sample stars (Fig.~\ref{fig1}(b) and Fig.~\ref{full_rb_mean}), most Li-rich stars (A(Li) $>$ 2.7 dex) younger than 4 Gyr originate from the outer disc ($\sim$70\% of these stars have R$_{\rm birth}$ $>$ 10 kpc). These stars exhibit an increasing trend in Li abundance over time, with a significant number exceeding the primordial cosmological Li abundance. Conversely, stars older than 6 Gyr predominantly originate from the inner disc. These older stars also demonstrate an increase in Li abundance over time, although this trend is relatively flatter compared with that of younger stars. Interestingly, the intermediate-age stars with age between 4.5 Gyr and 6 Gyr exhibit a decreasing trend in Li abundance over time. As illustrated in Fig.~\ref{fig1}(c), most of these stars have $T_{\rm eff}$ below 6200 K. This $T_{\rm eff}$ range suggests that they may correspond to the Li-dip main-sequence turn-off stars and subgiants identified in \cite{2024MNRAS.528.5394W}. A detailed discussion of this phenomenon will be provided in Section \ref{sec:4.2}.

Fig.~\ref{fig:feh-ali} illustrates the [Fe/H]-A(Li) distributions of the sample stars. In Fig.~\ref{fig:feh-ali}(a), stars younger than 4 Gyr exhibit a narrow range of metallicity, primarily between $-$0.4 dex and +0.1 dex. In contrast, intermediate-age stars, between 4 Gyr and 8 Gyr, show a broader range of metallicity, spanning from $-$0.6 dex to +0.4 dex. 
Both young and intermediate-age stars exhibit an increase in Li abundance with metallicity, consistent with previous studies of the thin disc \citep[e.g.,][]{2016A&A...595A..18G,2018A&A...610A..38F,2018A&A...615A.151B}. Moreover, since stars with [Fe/H] $\gtrsim$ 0 are primarily intermediate-age stars, it appears that the Li abundance of thin disc stars (age $<$ 8 Gyr) decreases at super-solar metallicities \citep{2015A&A...576A..69D,2016A&A...595A..18G,2020A&A...634A.130B,2020AJ....159...90S}. This phenomenon arises because the stars with the highest metallicities in the thin disc are not the youngest in the sample (Fig.~\ref{fig:feh-ali}(a)). As shown in Fig.~\ref{fig:feh-ali}(b), these stars predominantly originate closer to the Galactic center \citep{2023MNRAS.520.4815Z}, while stars with lower metallicity are more likely to form in the solar neighborhood. Older stars (age $>$ 8 Gyr) have significantly lower metallicity, mostly below 0, with many stars below $-$0.4 dex. According to the age criteria from \cite{2018A&A...615A.151B}, these stars belong to the thick disc population. Their Li abundance decreases with metallicity (Fig.~\ref{fig:feh-ali-old}), consistent with the results in \cite{2018A&A...615A.151B}. Additionally, these thick disc stars mostly originate from the inner disc, with smaller birth radii (Fig.~\ref{fig:feh-ali}(b)).

\begin{figure}
\includegraphics[scale=1.1]{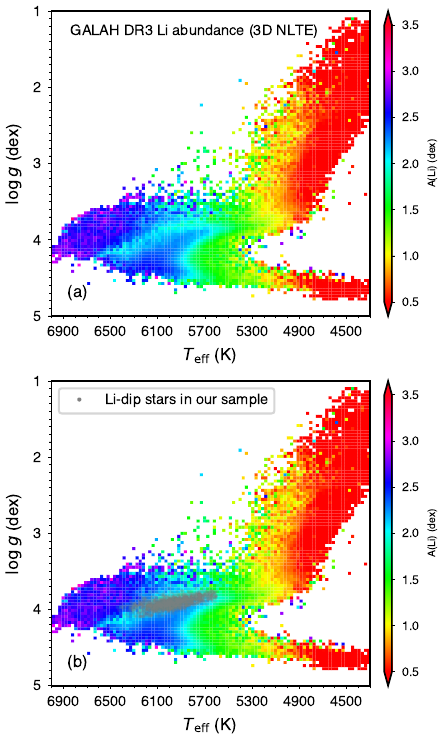}
\caption{Mean A(Li) binned in $T_{\rm eff}$ and $\log g$ for GALAH DR3 data (top), with the Li-dip stars (4-5 Gyr) found in our work overplotted (bottom).  \label{fig5}}
\end{figure}

\begin{figure}
\includegraphics[scale=1.1]{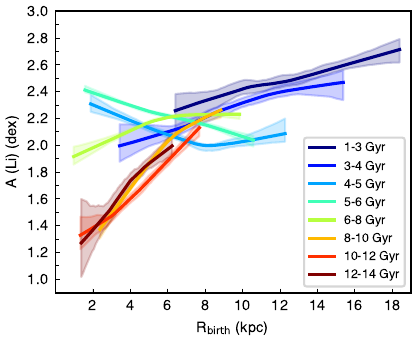}
% \plotone{Figure2/3D_Li_rg_gradient_0605_jet.png}
\caption{Radial Li abundance profile in bins of age, with respect to R$_{\rm birth}$.
Each line represents the local nonparametric regression fitting to the distribution of sample stars in this age bin. The shaded regions indicate the 95\%
confidence interval around the fitting result by performing bootstrap resampling.
\label{fig6}}
\end{figure}

\subsection{Age-A(Li) relations in each spatial bins} \label{sec:4.2}

In this section, we employ the guiding center radius R$_{\rm guide}$ and birth radius R$_{\rm birth}$ to investigate the variations in the age-A(Li) relations across different positions in the Milky Way. The guiding center radius reflects the average orbital radius of stars at present, allowing us to distinguish stars located in various regions of the Galactic disc. On the other hand, the birth radius indicates the orbital radius at the time of a star's formation, which enables us to trace the star's origin within the Galaxy.

We divide our sample stars into nine subsamples based on their guiding center radius R$_{\rm guide}$ and z$_{\rm max}$ (maximum vertical distance
from the disc plane) distribution, as shown in Fig.~\ref{fig:rg-zmax}. 
z$_{\rm max}$ utilised in this work come from the GALAH DR3 value-added catalogue (VAC) \citep{2021MNRAS.506..150B}, have been calculated using the Python package \texttt{Galpy} \citep{2015ApJS..216...29B}, with the details of assumed Milky Way potential and solar kinematic parameters presented in \citet{2021MNRAS.506..150B}.

We refer to three radial bins delimited at R$_{\rm guide}$ = 8 $\pm$ 1.00 kpc as the inner (R$_{\rm guide}$ $<$ 7 kpc), local (7 kpc $<$ R$_{\rm guide}$ $<$ 9 kpc), and outer (R$_{\rm guide}$ $>$ 9 kpc) regions of the disc. Each of the radial bins is divided into a low-z$_{\rm max}$ (z$_{\rm max}$ $<$ 0.3 kpc) bin, a intermediate-z$_{\rm max}$ (0.3 kpc $<$ z$_{\rm max}$ $<$ 0.7 kpc) bin, and a high-z$_{\rm max}$ (z$_{\rm max}$ $>$ 0.7 kpc) bin at z$_{\rm max}$ = 0.3 kpc and 0.7 kpc as shown in Fig.~\ref{9fig}.
Except for the sample stars shown in Fig.~\ref{9fig}(a), nearly all spatially divided subsamples exhibit a decreasing trend in Li abundance from $\sim$6 Gyr to $\sim$4 Gyr, with the minimum Li abundance occurring around 4.5 Gyr.
This suggests that the age range of Li-dip stars is 4-5 Gyr, irrespective of positions in the Milky Way. Additionally, aside from the stars in Fig.~\ref{9fig}(a,b,d), the remaining subsamples demonstrate a rapid increase in Li abundance after 4 Gyr. 

To investigate the origins of Li-dip stars and young Li-rich stars, we represent each star's birth radius by colour in Fig.~\ref{9fig-color}. Our results indicate that young Li-rich stars generally have the largest birth radii, mostly exceeding 12 kpc, implying migration from the outer disc. In contrast, the Li-dip stars identified in Fig.~\ref{9fig} have birth radii near the Sun, indicating minor influence from radial migration. 

Furthermore, Fig.~\ref{9fig-color}(a-c) reveals an intriguing pattern: stars born in the inner disc (orange points) exhibit significant Li enrichment from $\sim$8 Gyr to $\sim$6 Gyr, with this enhancement being most pronounced in the high-z$_{\rm max}$ region of the inner disc. This rapid Li enrichment trend parallels that observed in young stars (age $<$ 4 Gyr), suggesting that the inner disc experienced a similar enrichment process as the outer disc. This Li enrichment of inner disc stars is also observable in the local disc (Fig.~\ref{9fig-color}(d-f)), although their trend may be obscured by the presence of local disc stars. These findings highlight the importance of stellar birth radius in understanding the Li enrichment history in the Milky Way. The observed Li abundance trends in the solar neighbourhood may be influenced by stars originating from different regions of the Galactic disc. Before this period (8 Gyr to 6 Gyr) of Li enrichment, inner disc stars at high-z$_{\rm max}$ region (Fig.~\ref{9fig}(a) and Fig.~\ref{9fig-color}(a)) experienced Li depletion from 14 Gyr to 10 Gyr, with Li abundance decreasing by $\sim$0.5 dex.

To verify whether the Li-dip stars in our sample are located in the Li-dip region identified in \cite{2024MNRAS.528.5394W}, we select sample stars within the age range of 4-5 Gyr and plot them on the Kiel diagram. Fig.~\ref{fig5}(a) displays the distribution of GALAH DR3 3D NLTE Li abundance on the Kiel diagram, showing that Li-dip stars extend from $T_{\rm eff}$ $\approx$ 6500 K, $\log g$ $\approx$ 4.2 dex to subgiants at $T_{\rm eff}$ $\approx$ 5700 K and $\log g$ $\approx$ 3.8 dex. In Fig.~\ref{fig5}(b), the Li-dip stars identified in our sample (grey points) align closely with the Li-dip region, extending from main sequence turnoff stars to sub-giants. Therefore, we suggest that the age range of Li-dip MSTO and subginat stars, based on 3D NLTE Li abundances and our reliable stellar age, is 4-5 Gyr.

\subsection{Time evolution of A(Li) gradients}

To better understand how the Li abundance in the Galactic disc varies with time, we now focus on the time evolution of Li abundance gradients. We adopt the birth position (R$_{\rm birth}$) of stars as the radial distance scale for studying ISM evolution, considering it a more rational choice compared with the current position, which inherently incorporates the effects of radial migration. 

Fig.~\ref{fig6} depicts the temporal evolution of A(Li) gradients in terms of R$_{\rm birth}$, revealing a complex trend. This evolution can be broadly categorised into three distinct epochs: 14-6 Gyr ago, 6-4 Gyr ago, and 4-1 Gyr ago.
In the first period, spanning from 14 Gyr ago to 6 Gyr ago, the Li abundance profile demonstrates a positive gradient. Specifically, from 14 Gyr ago to 10 Gyr ago, the Li abundance in the inner disc exhibits a slight decline. This trend reverses from 10 Gyr ago to 6 Gyr ago, with the Li abundance in the inner disc increasing, accompanied by a gradual flattening of the gradient.
The second period, from 6 Gyr ago to 4 Gyr ago, is characterised by a transition in the Li abundance profile from a negative gradient to a broken gradient. This phase is marked by a decrease in Li abundance within the inner disc and local disc, corresponding to the age range of Li-dip stars. The observed broken gradient is attributed to the increase in Li abundance within the outer disc.
During the third period, from 4 Gyr ago to 1 Gyr ago, the Li abundance profile reverts to a positive gradient. This phase is distinguished by a gradual increase in Li abundance in both the local disc and outer disc.

The variation in the Li abundance gradient reflects the inside-out formation of the Galactic disc \citep[e.g.,][]{2019ApJ...884...99F}, wherein the disc expands from the inner to the outer regions. Over time, the overall Li abundance increases, except for the depletion observed in old (10-14 Gyr) stars and the Li-dip in intermediate-age (4-5 Gyr) stars. Throughout other epochs, the Li abundance in the Galactic disc consistently rises.

\section{Conclusions} \label{sec: conclu}

Utilising a sample of 22,034 main-sequence turn-off stars and subgiants from \cite{2023arXiv231105815S}, we investigated temporal and spatial variations in Li abundance within the Milky Way. Our study incorporates precise stellar ages, 3D NLTE Li abundances, and birth radii for the selected star samples.

We find that the temporal evolution of Li abundance in our sample is complex. Initially, there is a gradual increase in Li abundance from 14 Gyr to 6 Gyr, followed by a decline between 6 Gyr and 4.5 Gyr, and subsequently a rapid increase after 4.5 Gyr. Notably, young Li-rich stars with ages less than 4 Gyr primarily originate from the outer disc.

By binning the sample according to the guiding center radius and z$_{\rm max}$, our results indicate that young Li-rich stars in all spatial regions originate from the outer disc and have migrated radially to the local and inner discs. 
The stars originating from the inner disc experienced a rapid Li enrichment process from 8 Gyr ago to 6 Gyr ago, analogous to the enrichment observed in the outer disc stars after 4 Gyr ago.
Additionally, we find that the age range of Li-dip stars is 4-5 Gyr, encompassing evolution stages from the main-sequence turn-off stars to subgiants. 

Furthermore, our investigation into the temporal evolution of Li abundance gradients reveals three distinct periods: 14-6 Gyr ago, 6-4 Gyr ago, and 4-1 Gyr ago. Initially, the Li abundance gradient is positive. During the second epoch, it transitions to a negative and broken gradient, primarily influenced by the Li-dip stars. In the final epoch, the gradient reverts to a positive trend.

Our findings highlight the crucial roles of stellar age and birth radius in understanding the complex patterns of Li abundance variations, which are affected by radial migration.

\section*{Acknowledgements}

This work used the data from the GALAH survey, which is based on observations made at the Anglo Australian Telescope, under programs A/2013B/13, A/2014A/25, A/2015A/19, A/2017A/18, and 2020B/23.
This work has made use of data from the European Space Agency (ESA) mission Gaia (\url{https://www.cosmos.esa.int/gaia}), processed by the Gaia Data Processing and Analysis Consortium (DPAC, \url{https://www.cosmos.esa.int/web/gaia/dpac/consortium}). Funding for the DPAC has been provided by national institutions, in particular the institutions participating in the Gaia Multilateral Agreement.
This work is supported by the Joint Research Fund in Astronomy (U2031203) under cooperative agreement between the National Natural Science Foundation of China (NSFC) and Chinese Academy of Sciences (CAS), the NSFC grants (12090040, 12090042, 12373020, 12403037), and the National Key R$\&$D Program of China No. 2019YFA0405503, 2023YFE0107800. This work is partially supported by the Scholar Program of Beijing Academy of Science and Technology (DZ:BS202002).

%%%%%%%%%%%%%%%%%%%%%%%%%%%%%%%%%%%%%%%%%%%%%%%%%%
\section*{Data Availability}

The data underlying this article will be shared on reasonable request to the corresponding author.

%%%%%%%%%%%%%%%%%%%% REFERENCES %%%%%%%%%%%%%%%%%%

% The best way to enter references is to use BibTeX:

\bibliographystyle{mnras}
\bibliography{ref} % if your bibtex file is called example.bib

% Alternatively you could enter them by hand, like this:
% This method is tedious and prone to error if you have lots of references
% \begin{thebibliography}{99}
% \bibitem[\protect\citeauthoryear{Sun et al.}{2023}]{2023arXiv231105815S} Sun T., Bi S., Chen X., Chen Y., Liu C., Zhang X., Li T., et al., 2023, arXiv, arXiv:2311.05815. doi:10.48550/arXiv.2311.05815
% \bibitem[\protect\citeauthoryear{Author}{2012}]{Author2012}
% Author A.~N., 2013, Journal of Improbable Astronomy, 1, 1
% \bibitem[\protect\citeauthoryear{Others}{2013}]{Others2013}
% Others S., 2012, Journal of Interesting Stuff, 17, 198
% \end{thebibliography}

%%%%%%%%%%%%%%%%%%%%%%%%%%%%%%%%%%%%%%%%%%%%%%%%%%

%%%%%%%%%%%%%%%%% APPENDICES %%%%%%%%%%%%%%%%%%%%%

\appendix 

\section{Distribution of Li Abundance and Birth Radius}
\label{appendix}

\begin{figure}
\includegraphics[scale=0.44]{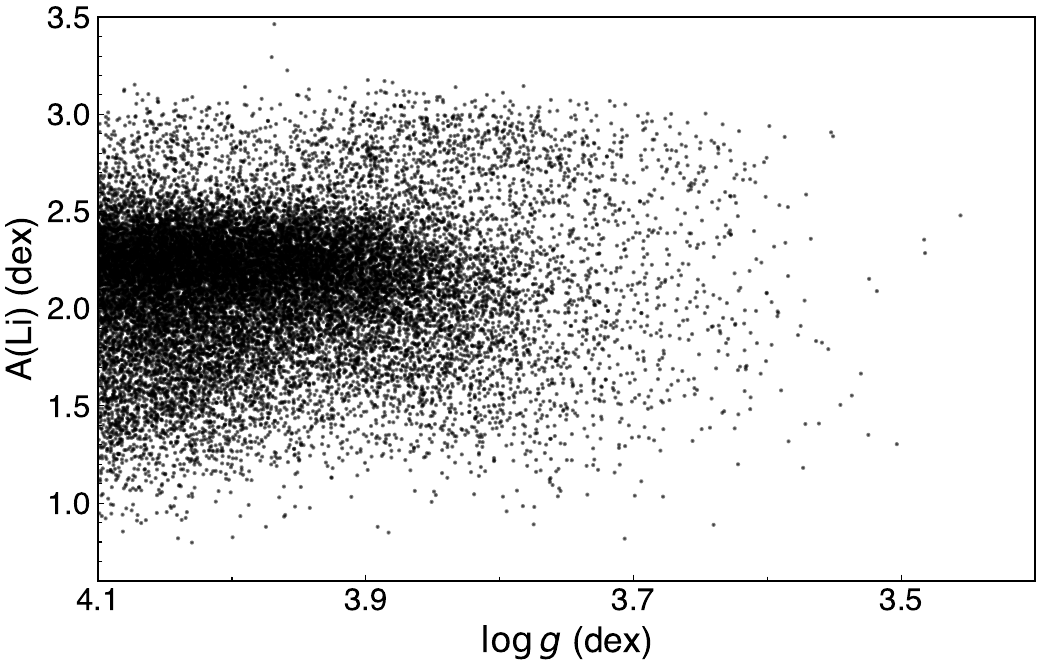}
\caption{$\log g$-A(Li) distribution of our sample stars.
\label{logg_ali}}
\end{figure}

\begin{figure}
\includegraphics[scale=1.15]{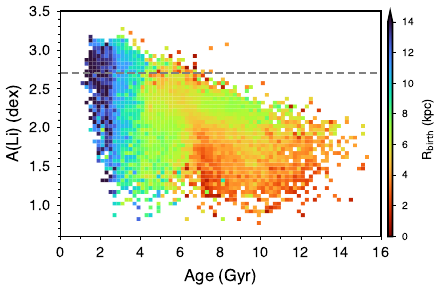}
\caption{Mean R$_{\rm birth}$ binned in age and A(Li) for sample stars. 
\label{full_rb_mean}}
\end{figure}

\section{[Fe/H]-A(Li) distributions of old stars}
\label{appendix2}

\begin{figure}
\includegraphics[scale=0.62]{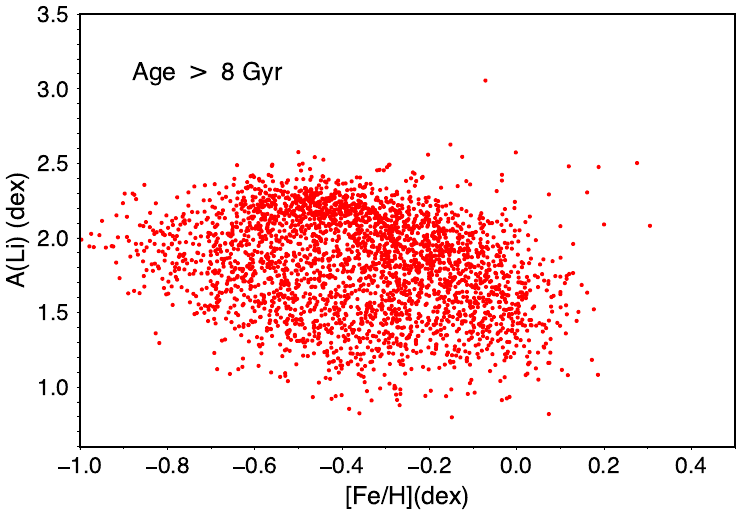}
\caption{[Fe/H]-A(Li) distribution of old stars with age $>$ 8 Gyr.
\label{fig:feh-ali-old}}
\end{figure}

%%%%%%%%%%%%%%%%%%%%%%%%%%%%%%%%%%%%%%%%%%%%%%%%%%

% Don't change these lines
\bsp	% typesetting comment
\label{lastpage}
\end{document}